\documentstyle[aps,prl,graphicx]{revtex}

\input psfig.sty

\begin{document}
\twocolumn 
\wideabs{  
\title{Using surface-wave spectroscopy to characterize tilt modes of a vortex in a Bose-Einstein Condensate}
\author{P.~C. Haljan, B.~P. Anderson\cite{qpdNIST}, I. Coddington, and E.~A. Cornell\cite{qpdNIST}}
\address{JILA, National Institute of Standards and Technology and Department of Physics, \\
University of Colorado, Boulder, Colorado 80309-0440}
\date{\today}

\maketitle

\begin{abstract}
A vortex in a condensate in a nonspherical trapping potential
will in general experience a torque. The torque will induce
tilting of the direction of the vortex axis. We observe this
behavior experimentally and show that by applying small
distortions to the trapping potential, we can control the
tilting behaviour. By suppressing vortex tilt, we have been able
to hold the vortex axis along the line of sight for up to 15
seconds. Alternatively, we can induce a 180$^{\circ}$ tilt,
effectively reversing the charge on the vortex as observed in
the lab frame. We characterize the vortex non-destructively with
a surface-wave spectroscopic technique.
\\ 

PACS number(s):03.75.Fi, 67.90.+z, 67.57.Fg, 32.80.Pj 

\end{abstract}
 } 
\par
The decay of ``persistent" supercurrents, be they in superfluids or superconductors, is intimately connected to the dynamical behavior of vortices. A magnetically trapped, gas-phase Bose-Einstein condensate (BEC) provides a useful laboratory for characterizing the microscopic behavior of individual vortices subject to various controlled perturbations \cite{Matthews1999b,Madison2000a,Anderson2000a,Chevy2000a}. In one recent experiment, a vortex core in a near-spherical condensate ``vanished" \cite{Anderson2000a} from view apparently without moving out to the edge of the sample and annihilating there. Another group \cite{Madison2000a} found that the empirical critical rotation velocity for the formation of a vortex in an elongated condensate is much higher than can be accounted for by a simple model of the vortex as a rigid line-defect without any dynamics along its length. Feder \textit{et al.} \cite{Feder2000} explain the latter observation by showing that the higher rotation rate is necessary to suppress the growth of anomalous normal modes (``bending" modes) of the vortex. In this paper we study the lowest odd-order normal mode, which in our near-spherical geometry corresponds not to a bend but to a tilting of the vortex orientation. We show that the ``vanishing" vortex of ref. \cite{Anderson2000a} was in fact due to tilting of the vortex away from the line of sight. Such uncontrolled tilting was caused by residual asphericity in the condensate's confining potential. By tailoring the asphericity, we have learned to control the tilting dynamics.
\par
The tilting behaviour of a vortex in a condensate, confined in a slightly aspheric, parabolic potential, is discussed in detail by Svidzinsky and Fetter \cite{tilttheory}, starting from the Gross-Pitaevskii equation. Only a brief summary of their theoretical results is given here. The vortex direction satisfies a set of equations similar in form to the Euler equations for rigid body rotation, familiar from classical mechanics. As a result, the tilting dynamics of a quantized vortex in a confined BEC are reminiscent of the behaviour of a freely spinning rigid body as seen in the body-fixed frame \cite{FetterandWalecka}. In the case of the spinning rigid body, there are two stable axes about which precession of the spin direction will occur, namely the axes with the largest and smallest moments of inertia. The intermediate axis is unstable and no precession occurs about that direction. If the spin direction lies initially near the intermediate axis, it will evolve away from its initial orientation. Similarly, for a quantized vortex in a BEC, tilt precession of the vortex is predicted to occur about two stable axes, given in this case by the tight and weak directions of confinement. The unstable axis corresponds to the direction of intermediate trapping strength.
\par
The tilting dynamics of a vortex are constrained by two integrals of motion, one in particular corresponding to conservation of energy. Physically, this implies that, as the vortex tilts, its direction follows an angular trajectory which is a contour of constant vortex length, or equivalently constant energy. As is the case for all the normal modes of a vortex, the precession frequency for a vortex tilting about a stable axis is predicted to scale with the rate of vortex fluid circulation evaluated at the condensate edge. For our typical conditions, this fluid rotation is near 0.3 Hz. The theoretical frequency for tilt precession depends additionally on the magnitude and character of the trap asymmetries and the initial orientation of the vortex.

\par
In our experiment, we first use a wavefunction engineering
technique to make singly quantized vortices in a two-component
BEC \cite{Matthews1999b,Williamstheory}. The two components,
which are two different hyperfine levels of $^{87}$Rb, are
magnetically confined together in a nominally spherical,
harmonic TOP trap \cite{Ensher1998a}, parameterized by a
trapping frequency $\omega_{trap}/2\pi$ = 7.8(1) Hz. The vortex
formation process leaves one component in the circulating state
while the other component, which is non-rotating, fills out the
core. The vortex is formed initially aligned along the line of
sight. We take a non-destructive picture of the two-component
vortex to record the initial displacement of the core with
respect to the center of the condensate cloud. We then create a
bare vortex by selectively removing the fluid filling the core \cite{Anderson2000a}, which shrinks down to a size below our 
imaging resolution.
\par
This paper deals exclusively with the dynamics of bare vortices
in a single-component BEC. The bare vortex state is formed with
about $2\cdot10^5$ atoms at a temperature of $T/T_{c}=0.8(1)$
where the critical temperature $T_{c}$ is 20(6) nK. The number
of atoms in the condensate is determined (in the Thomas-Fermi
limit) from the condensate radius R, equal to 20.8(4) $\mu$m on
average. Following the creation process and core removal, the
condensate is held for a variable holding time in its confining
potential, and then probed for the presence of the vortex.
\par
We have previously reported that the visibility of vortex cores
was lost after a holding time of about 1 s in a nominally
spherical trap \cite{Anderson2000a}. In those experiments, the
confining potential was suddenly removed, and the condensate
allowed to expand ballistically before imaging. The presence of
a vortex was detected in the expansion image as a dimple in the
condensate's density distribution. The topological nature of a
quantized vortex means that the only way for a BEC to rid itself
of a vortex is for the core to make its way to the edge of the
cloud and annihilate there; the vortex cannot gradually spin
down. Our vortices, however, seemed to disappear without any
visible radial motion of the cores outward. We did notice that
the contrast of the cores as observed in expansion decreased at
longer holding times before being lost altogether. These results
could be explained by a tilting of the vortex away from the line
of sight. Because the vortex core is such a narrow feature, only
small deviations from the line of sight ($\sim 20^{\circ}$) are
necessary for contrast to be lost below the noise threshold in
the expansion images.
\par
In order to circumvent this imaging limitation, we have
implemented an alternative method of vortex detection that is
more robust against tilting of the vortex from the line of
sight, and that has the additional advantage of being sensitive
to the handedness of the vortex circulation. This detection
technique uses the idea that the collective excitation
frequencies of a trapped condensate are sensitive to the
presence of a vortex
\cite{quaddetection}. We make use of quadrupolar surface-wave excitations, that is, surface-wave excitations which carry angular momentum $l=2$. (The angular momentum of these perturbative excitations should not be confused with the angular momentum per particle of the bulk of the condensate, which in the presence of a well-centered vortex approaches $l=1$). The projection of angular momentum of an $l=2$ surface-wave excitation onto the axis of the vortex direction can have components $m=0,\pm1,\pm2$. In the absence of a vortex, the counter-propagating $\pm m$ modes are degenerate, but in the presence of a vortex the handedness of the fluid flow breaks time-reversal symmetry and lifts the degeneracy. The use of surface-wave spectroscopy for vortex detection was first demonstrated experimentally by Chevy \textit{et al.} \cite{Chevy2000a}.
\par
To detect a vortex, we excite an $l=2$, $m_{z}=0$ excitation about the vertical $\hat{z}$ axis by modulating the trapping potential for a single cycle. Along the line of sight, the $\hat{x}$

\begin{figure}  
\begin{center}
\psfig{figure=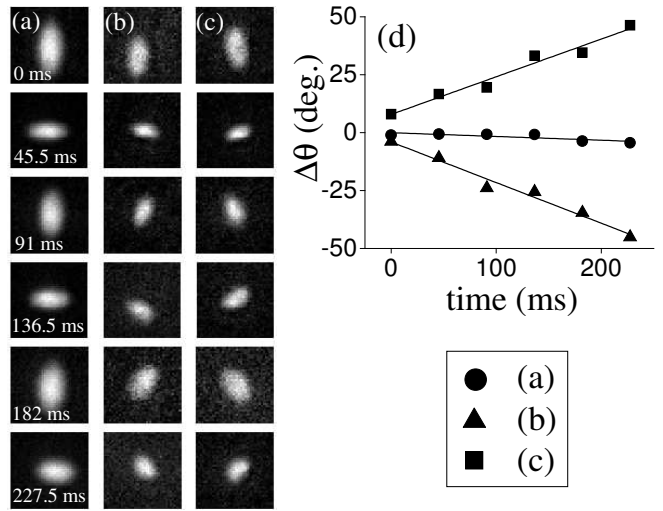,width=1\linewidth,clip=}
\end{center}
\caption {Using surface excitations for  \textit{in situ} detection of a
vortex in a confined BEC. (a), (b) and (c) are each a series of
non-destructive images of the quadrupolar mode, after
excitation. The pictures are strobed at 45.5 ms, half the
excitation period. (a) is the case of a vortex free condensate;
(b) and (c) show the excitations in the presence of a vortex
whose core is normal to the plane of the page. The vortices in
(b) and (c) have opposite handedness. The principal axes of the
ellipse-shaped quadrupolar mode precess in the direction of the
fluid flow. The images are each fit to an elliptical
distribution with orientation $\Delta\theta$ of the principal
axes. The orientation, expressed as an angular deviation from
the vertical and horizontal axes, is plotted versus time in (d)
for each of the cases (a), (b) and (c). A linear fit has been
applied to the data to determine a precession frequency of the
principal axes, -0.49(4) Hz for (b) and 0.45(5) Hz for
(c).}
\label{quaddetection}
\end{figure}

\noindent
axis, the $m_{z}=0$ mode projects onto a superposition of $m_{x}=+2$, $m_{x}=-2$, and $m_{x}=0$ modes. The $m_{x}=+2$ and $m_{x}=-2$ superposition may be thought of as a standing-wave of clockwise and counterclockwise surface waves. Along the line of sight we observe the cloud alternately stretch along two principal axes, first vertical, then horizontal. The excitation period is measured to be 11.0(2) Hz in agreement with the predicted value $\sqrt{2}\omega_{trap}$ \cite{Stringari1996a}. The presence of a vortex along the line of sight induces a splitting of the clockwise and counterclockwise wave velocities, so that the nodes of the standing-wave are not completely fixed. As a result we observe a precession of the principal axes of the quadrupolar excitation. We record the precession in a sequence of seven non-destructive images, strobed at half the quadrupolar excitation period. A typical data set is shown in Fig.~\ref{quaddetection}. Opposite vortex circulation clearly leads to opposite precession of the principal axes, as is evident in Figs.~\ref{quaddetection}(b) and (c).
\par
For a vortex whose direction is tilted with respect to the line of sight, we still expect precession of the quadrupolar principal axes to be induced about the vortex direction. The main limitation to vortex detection in this case is that there must be sufficient projection of the quadrupolar precession onto the imaging camera. In the extreme case where the vortex is oriented at 90$^{\circ}$ to the line of sight, any quadrupolar precession will also be about an axis perpendicular to the line of sight, in other words, not readily observable on the imaging camera.
\par
Armed with our new detection method, we revisited the topic of
longevity of vortices in our nominally spherical trap. The
precession of surface waves clearly revealed the presence of
vortices at 1.5 s and at 2 s holding time, well after the
vortices ceased to be visible in the expansion images
\cite{Anderson2000a}. Given that the surface-wave probe for
vortices should be relatively robust to small tilts of the
vortices with respect to the line of sight, we came to the
tentative conclusion that our vortices were in fact tilting away
from the line of sight during the first few seconds after their
creation. This behavior could be accounted for by small
($\leq$3$\%$) residual asymmetries in our trapping potential.
\par
Further study of the tilting required better characterization of
the trap potential, and we achieved this by deliberately
introducing tailored deformations to the confining potential,
deformations that were certain to overwhelm the residual
asymmetry. The results of three such experiments are presented
here: first, the suppression of vortex tilt; second, the
generation and observation of vortex tilt about a stable axis;
and, third, the manipulation of a vortex's orientation through
controlled deformations 

\begin{figure}  
\begin{center}
\psfig{figure=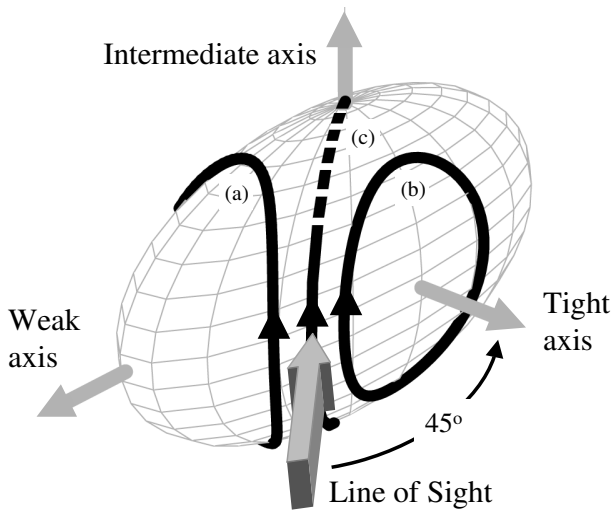,width=1\linewidth,clip=}
\end{center}
\caption {Possible tilting behaviour of a vortex in a confined
BEC is illustrated for the conditions of the second tilting
experiment (see text). The axes of symmetry for the triaxial
confining potential are indicated. Three possible trajectories
for a tilting vortex, initially aligned near the line of sight,
are traced out on the surface of the condensate. Depending on
the initial orientation of the vortex and the relative values of
the confinement strengths, the vortex direction could (a)
precess about the weak axis, (b) precess about the tight axis,
or (c) drift towards the intermediate axis along a `saddle'
line.}
\label{sketchrevival}
\end{figure}

\noindent
of the BEC's confining potential.
\par
In the first experiment, the spherical confining potential is
squeezed by 10$\%$ along the line of sight following the
creation of a vortex in the BEC \cite{holeofdeath,squeezetime}.
This places the vortex along the tight, stable axis of the trap.
As a result, tilting of the vortex away from the line of sight
should be suppressed. After a variable holding time in the
squeezed trap, the confining potential is returned to spherical
symmetry and the condensate is then quickly probed for the presence
of the vortex. Using both the surface-wave and expansion
techniques, we have detected vortices lasting for holding times
up to 15 s. This is a significant improvement over our
previously published limit of 1 s. Although visibility has been
restored at long times by holding the condensates in a squeezed
trap, it is important to note that only well-centered vortices
have been found to survive to the longest times. This is
consistent with a model of initially offset cores spiraling out
of the condensate under the influence of thermal damping
\cite{dampingrefs}.
\par
\begin{figure} 
\begin{center}
\psfig{figure=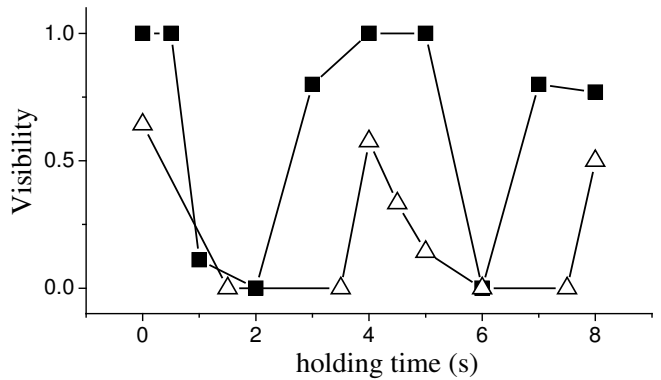,width=1\linewidth,clip=}
\end{center}
\caption{Visibility of a vortex in a BEC is plotted versus the time the BEC is held in a trap squeezed along an axis at $45^{\circ}$ to the line of sight. Visibility is the probability of detecting a vortex over several shots. Filled squares indicate vortex detection with the surface-wave technique; open triangles indicate detection with the expansion technique.}
\label{revivaldata}
\end{figure}

The second experiment proceeds in the same way as the first except the spherical trap is now squeezed along a horizontal axis at 45$^{\circ}$ to the line of sight. The squeezing process produces a triaxial confining potential, where the tight and weak directions of confinement lie in the horizontal plane while the intermediate axis is vertical. The measured trap asymmetries together with the initial vortex direction indicate that the vortex should precess about the weak axis, although slightly different conditions could lead to rather different behavior (see Fig.~\ref{sketchrevival} \cite{trajectories}). In any case, as the vortex tilts away from the line of sight to a maximum excursion of 90$^{\circ}$ from the imaging axis, the visibility of the vortex will gradually disappear. Eventually, if the vortex precesses about the stable axis and back into the line of sight, it should become visible again in a `revival of visibility'. This effect is seen in Fig.~\ref{revivaldata}, where vortex visibility is plotted as a function of 

\noindent
holding time in the squeezed trap.
Visibility is defined as the probability of detecting a vortex
at a given time and is obtained from a set of approximately ten
shots for each point plotted. Vortices whose initial offset from
trap center exceeds a prescribed value of $\sim$0.15 R are
omitted from the analysis. This is done to reduce blurring of
the revival due to possible position-dependent tilt rates, and,
moreover, to prevent loss of contrast at long times due to the
drift of cores out of the condensate. The vortex detection is
accomplished with both surface-wave and expansion techniques.
The revival measured with expansion imaging is much more
narrowly resolved in time, as is to be expected from this
method's greater sensitivity to alignment of a vortex with the
line of sight.
\par
The frequency of the tilt mode, determined from the data in
Fig.~\ref{revivaldata}, is 0.25(2) Hz. The theory of Svidzinsky
and Fetter \cite{tilttheory} predicts the lowest-order
odd-parity vortex mode should have a frequency below 0.3 Hz for
our values of the Thomas-Fermi radius, confinement asymmetry,
and initial vortex angle. Uncertainties in the confinement
asymmetry preclude a more precise determination of the
theoretical tilting frequency. The uncertainties are such that
the vortex could lie on an initial trajectory arbitrarily close
to a `saddle' line, where the revival time should diverge (see
Fig.~\ref{sketchrevival}). The high contrast of the two revivals
in Fig.~\ref{revivaldata} clearly indicates, however, that the shot-to-shot fluctuations of the confinement asymmetry and of the initial position of the vortex are small.
\par
The third and final tilting experiment demonstrates the reversal
of vortex handedness. The vortex, as usual, 

\begin{figure}     
\begin{center}
\psfig{figure=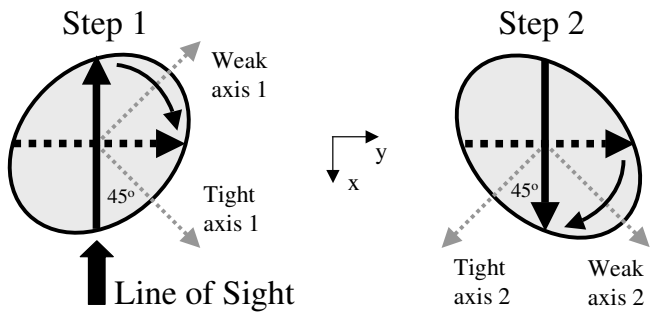,width=1\linewidth,clip=}
\end{center}
\caption {A sequence of trap deformations to reverse the direction of a
vortex is shown. The figure shows a horizontal crosssection of
the BEC as seen from above. The vortex is initially formed along
the line of sight and aligned with the -$\hat{x}$ axis. In step
1, the trap is squeezed in the horizontal plane along tight axis
1. The vortex is then allowed to precess for 1.5 s, at which
point it should be aligned with the $\hat{y}$ axis. The new
direction of the vortex is indicated by a dotted arrow. In step
2, the trap is squeezed along tight axis 2 and the vortex
allowed to precess another 1.5 s until it is aligned with the
$\hat{x}$ axis. To an observer looking down the line of sight,
the vortex has effectively reversed its handedness. Changing the
time allowed for vortex precession by 0.5 s inhibited the
effect. For the correct timing, the flipping process was found
to be successful 80$\%$ of the time.}
\label{sketchflip}
\end{figure}

\noindent
is formed along the line of sight. The direction of the vortex is then flipped 180$^{\circ}$ by deforming the trapping potential in two steps, as illustrated in Fig.~\ref{sketchflip}. From the perspective of the line of sight, the `lab frame,' the vortex has effectively flipped its handedness. Surface-wave detection is used to verify this, with pictures similar to Figs.~\ref{quaddetection}(b) and (c) being obtained before and after the flipping process respectively. Manipulation of vortex direction may prove to be a useful technique in a TOP trap. For example, maneuvering a vortex to an arbitrary direction followed by a second stage of wavefunction engineering could produce a condensate wavefunction with more complicated, three-dimensional topological structure.
\par
In conclusion, we have described a non-destructive method of
vortex detection and used it to characterize the tilt modes of a
bare vortex in a trapped BEC. Control over the tilt of a vortex
has been demonstrated, including the suppression of tilting altogether.  The suppression of tilting, which maintains the
visibility of vortices out to long times, now permits the study
of the lifetime of vortices at finite temperature.
\par
We would like thank A.~L. Fetter, A.~A. Svidzinsky, and C.~E.
Wieman for helpful conversations. This work was supported by
funding from NSF, ONR, and NIST. P.C.~Haljan wishes to
acknowledge support from an NSERC fellowship.


%


\begin{thebibliography}{10}
\bibitem{holeofdeath}
A 10$\%$ tightening of the spherical trap is achieved by turning
off the component of the rotating TOP bias field along the
direction of the imaging axis. This leaves an oscillating bias
field perpendicular to the line of sight. The BEC equilibrium
position is sagged so far vertically that the oscillating
quadrupole zero of the squeezed TOP trap does not pass through
the cloud.
\bibitem{squeezetime}
The confining potential is squeezed over 200 ms, slow enough to avoid inducing large-amplitude shape oscillations of the BEC but fast enough so as not to interfere significantly with the tilting dynamics.
\end{thebibliography}
\end{document}